\documentclass[a4paper,11pt]{article}
\pdfoutput=1
\usepackage{jcappub}
\usepackage[british]{babel}
\usepackage{latexsym,amsmath,amssymb}
\usepackage{graphicx}
\usepackage{booktabs}
\usepackage{slashed}
\usepackage{bm}
\usepackage{epsfig}
\usepackage{dcolumn}

\newcommand{\nn}{\nonumber}
\newcommand{\be}{\begin{eqnarray}}
\newcommand{\ee}{\end{eqnarray}}

\newcommand{\rar}{\rightarrow}
\newcommand{\Rar}{\Rightarrow}

\def\dd{ \,\text{d} }
\def\d{\partial}
\def\+{\dagger}

\newcommand{\kpc}{\text{kpc}}
\newcommand{\Mpc}{\text{Mpc}}
\newcommand{\Gauss}{\text{Gauss}}

\newcommand{\cph}{\varphi}

\newcommand{\cH}{{\cal H}}
\newcommand{\cA}{{\cal A}}

\newcommand{\cP}{{\cal P}}
\newcommand{\order}[1]{{\cal O}\left(#1\right)}
\renewcommand{\v}[1]{{\boldsymbol{#1}}}
\newcommand{\vk}{\v{k}}
\newcommand{\vx}{\v{x}}

\title{Occupy magnetogenesis}

\author[a,b]{Peter G.~Tinyakov}
\author[b]{and Federico R.~Urban}

\affiliation[a]{Institute for Nuclear Research of the Russian Academy of Sciences, 117312, Moscow, Russia}
\affiliation[b]{Service de Physique Th\'eorique, Universit\'e Libre de Bruxelles, CP225, Boulevard du Triomphe, B-1050 Brussels, Belgium}

\emailAdd{petr.tiniakov@ulb.ac.be}
\emailAdd{furban@ulb.ac.be}

\abstract{The occupation number of a field is a sound discriminator between classical and quantum regimes.  In this pamphlet we give an overview of what we can learn about inflationary magnetogenesis just by looking at the occupation numbers of the classical magnetic fields observed today, and those of the quantum electromagnetic field during inflation.  The sole occupation number behaviour dictates that for a pure Maxwell theory in de Sitter space it appears impossible to match the two epochs.}


\notoc

\begin{document}
\maketitle

\newpage

\begin{flushright}
\emph{Hang your collar up inside\\
Hang your dollar on me\\
Listen to the water still\\
Listen to the causeway\\
You are mad and educated\\
Primitive and wild\\
Welcome to the occupation\\}
R.E.M., Welcome to the occupation
\end{flushright}

\section{Welcome to the occupation}\label{sec:wel}

\paragraph*{Scope.}
Our Universe is magnetised.  Observations present substantial evidence of this fact which holds true for a very wide range of lengthscales and structures, going from planets to clusters of galaxies, and likely beyond.  Compatibility with known features of the early Universe, primarily the cosmic microwave background and big bang nucleosynthesis, and late Universe, in the guise of rotation measures of distant objects, also bound several regions of the current magnetic field spectrum.  Excellent and detailed reviews on the subject report all essential numbers as well as comprehensive surveys of the proposals for the generation of cosmic magnetic fields, see for instance~\cite{Kronberg:1993vk,Grasso:2000wj,Kulsrud:2007an,Kandus:2010nw,Widrow:2011hs,Durrer:2013pga}.

The mystery is as of yet unsolved: the challenge is on.  The aim of this brief work is to clarify some aspects of magnetogenesis to provide insight on the very basic features a model is demanded to possess.  The focus here is on cosmological generation of magnetic seeds which can later evolve dynamically in the structures and topologies we detect.  The primary tool in this analysis is the \emph{occupation number} $n_\vk$ which determines the realm (quantum, classical) in which computations should be performed.

The boundary between classical and quantum is a brumous and a befuddling one, see for example~\cite{Albrecht:1992kf} and references therein.  The occupation number will reveal itself to be an extremely useful and well-formed tool to set some necessary conditions on primordial magnetogenesis models: it does us the service of a Land of Toys' Talking Cricket.  We walk through the Big Bang following the Cricket's adventures with the hope that restating known, perhaps not universally accepted or acknowledged, results in a different perspective contributes to shrug off some doubts found in recent literature~\cite{Finelli:2007fr,Durrer:2009ii,Marozzi:2011da,Campanelli:2013mea,Durrer:2013xla,Campanelli:2013fqa}, and form a useful reference for future efforts in this field.

\paragraph*{Definitions.}
The quantum Hamiltonian of a field $\cph$ and conjugate variable $\pi$, in momentum $\vk$ space, is
\be
\cH_\vk \equiv \cH^0_\vk + \cH^I_\vk = \pi_\vk \pi_{-\vk} + \cph_\vk \cph_{-\vk} + \cH^I_\vk \, ,
\ee
where the interaction term $\cH^I_\vk$ contains all interactions of the $(\pi,\cph)$ pair, including those with other fields.  This Hamiltonian can be used to describe cosmological perturbations during inflation, as well as the two physical polarisations of the massless electromagnetic field.  The canonical creation and annihilation operators are defined by
\be
\cph_\vk \equiv \frac{1}{\sqrt{2}} \left( a_\vk + a^\+_\vk \right) \, ,\\
\pi_\vk \equiv -i \sqrt{\frac{k}{2}} \left( a_\vk - a^\+_\vk \right) \, ,
\ee
in terms of which we rewrite
\be
\cH^0_\vk = k \left( a^\+_\vk a_\vk + a^\+_{-\vk} a_{-\vk} + 1 \right) \equiv k \left( N_\vk + 1 \right) \, .
\ee
The occupation number operator tells how many particles fill a given state $|\Psi\rangle$.  The expectation value of the number operator $n_\vk \equiv \langle\Psi| N_\vk |\Psi\rangle$ is then the scalar occupation number proper.

One possible definition of a classical system is one that has a large occupation number $n_\vk \gg 1$.  When this is the case classical equations of motion can (should) be employed.  The occupation number itself in this case reduces to $n_\vk \approx \langle \cH_\vk \rangle / k = E_\vk/k$, with $E_\vk$ the energy stored in a given Fourier mode of the classical field\footnote{For the relation between occupation number and squeezing see the Appendix.}.

\paragraph*{The numbers.}
What are the occupation numbers of the fields observed today?  As an example one can take some representative observed values.  For instance a $\mu\Gauss$ field intensity stretched across a typical galaxy of a few kpc returns some $n_{1/\kpc} \approx 10^{90}$.  Similarly, the lower bound of $10^{-15}\Gauss$ coherent on a few Mpc-sized void has $n_{1/\Mpc} \approx 10^{84}$.  The conversion factor between magnetic field energy densities and distances is $\sqrt{1\Gauss} \approx 4\cdot10^{28}/\Mpc$.

The occupation numbers are elephantine.  Thus, it is ``beyond any reasonable doubt'' that observed magnetic fields are completely classical today.  If it is believed that these are the outcome of some quantum dynamics taking place much earlier in the Universe, the question is: when did the quantum-to-classical transition happen?

\begin{table}[h]
\caption{Occupation numbers today$^*$}
\centering
\begin{tabular}{|l|lll|}
\hline\hline
Where		& Size				& Strength				& $n_\vk$ \\
\hline &&&\\ [-1.5ex]
Galaxy		& $0.1\div10$ kpc	& $10\mu\Gauss$			& $10^{88\div96}$ \\ [0.5ex]
Cluster		& Mpc				& (n$\div\mu)\Gauss$	& $10^{102\div96}$ \\ [0.5ex]
Void		& Mpc				& $10^{-15}\Gauss$		& $10^{84}$ \\ [0.5ex]
Earth		& $10^4$km			& $1\Gauss$				& $10^{54}$ \\ [0.5ex]
LHC			& 14m				& $10^4\Gauss$			& $10^{37}$ \\ [0.5ex]
\hline
\end{tabular} \vspace{0.5ex} \\
$^*$\footnotesize{These values do not reflect current Occupy movements trends.}
\end{table}

The next two sections sandwich the evolution of the occupation number between late-time classical regime (Sec.~\ref{sec:class}) and early inflationary quantum regime (Sec.~\ref{sec:quant}).  Sec.~\ref{sec:end} reports the result in a streamlined fashion.

\section{Classical late-time dynamics}\label{sec:class}

\paragraph*{Classical MHD at large scales.}
Large occupation number concedes the use of classical equations for the study of the dynamics.  In the late Universe, and especially at large scales, this is studied with the help of \emph{magnetohydrodynamics}~\cite{Brandenburg:2004jv}.

Circular currents induce magnetic fields. This can easily be seen by taking the curl of Amp\`ere's circuital law and substituting the electric field using Faraday's law of induction. One arrives at a wave equation for the --- physical --- magnetic field, $\square \vec{B}_p = -\vec{\nabla}\! \times \! \vec{J_p}$, that is sourced by the curl of the current.

Now let us immerse electromagnetism in our Universe.  Since the standard electromagnetic field is conformally invariant, and since the Friedmann-Lema\^itre-Robertson-Walker metric is conformally flat, Minkowski-space evolution equations are directly applicable to \emph{comoving} electric $\vec E \equiv a^2 \vec{E}_p$ and magnetic fields $\vec B \equiv a^2 \vec{B}_p$, where $a$ is the scale factor of the Universe.  The dynamics is known once Ohm's law, $\vec J = \hat\sigma (\vec E + \vec u \! \times \! \vec B)$, where $\vec u$ is the bulk velocity of the plasma and $\hat\sigma$ is the comoving conductivity that has units of an inverse length, is specified.

The comoving conductivity is related to the usual one by $\hat\sigma\equiv a\sigma$.  This is the crucial quantity here, since it governs much of the evolution of the electromagnetic field at large scales.  A detailed study of the behaviour of $\hat\sigma$ in the late Universe can be found in~\cite{Hollenstein:2012mb}.

The action for the electromagnetic field is
\be\label{lag}
S[A_{\mu}] \equiv \int\dd^4x\, \sqrt{-g} \left[ - \frac{1}{4}F_{\mu\nu} F^{\mu\nu} + J_\mu A^\mu \right] \, .
\ee
Here $F_{\mu\nu} \equiv \partial_\mu A_\nu-\partial_\nu A_\mu$ is the EM field tensor.  For the current, the simplified Ohm's law, $J_\mu = \sigma E_\mu = \sigma (0,-A_i'/a)$, can be used in the plasma frame where $u_\mu=0$.  In Coulomb gauge, $\d^i \! A_i = 0 = A_0$, the Euler-Lagrange equation become
\be\label{eom}
A_i'' - \nabla^2 A_i + \hat \sigma A_i' = 0 \, ;
\ee
prime is conformal time $\eta$ derivative.  This equation is best dealt with in Fourier space $A_i(\vx,\eta) \rar \cA_i(\vk,\eta)$ where it reads
\be\label{modes}
\cA_i'' + \hat\sigma \cA_i' + k^2 \cA_i = 0 \, .
\ee

Assuming the friction term to be time-independent for simplicity, the solution writes as
\be\label{cA}
\cA_i \approx C_1\, e^{-\frac12 \left\{ 1+\sqrt{1-4\kappa^2} \right\} \hat\sigma\eta} + C_2\, e^{-\frac12 \left\{ 1-\sqrt{1-4\kappa^2} \right\} \hat\sigma\eta} \, ,
\ee
where $\kappa \equiv k/\hat\sigma$.  The behaviour of the solution can be separated in the two regimes, $\kappa \ll 1$ and $\kappa \gg 1$, hence
\be
& \cA_i \approx e^{-\hat\sigma\eta/2} \left(C_1\, e^{-ik\eta} + C_2\, e^{ik\eta}\right) \qquad & \text{for} \quad \kappa \gg 1 \label{cAhighk} \, , \\
& \cA_i \approx C_1\, e^{-\hat\sigma\eta} + C_2\, e^{-\kappa^2\hat\sigma\eta} \qquad & \text{for} \quad \kappa \ll 1\label{cAlowk} \, .
\ee
The known result that the vector potential is either essentially frozen in the plasma or an oscillating and decaying wave, respectively, is recovered. The transition between these two regimes is given by the comoving conductivity of the plasma $\hat\sigma$.  In particular, for early times smaller than the inverse conductivity $\epsilon\equiv\hat\sigma\eta\ll1$:
\be
& \cA_i \sim e^{-\epsilon/2} \qquad & \text{for} \quad \kappa \gg 1 \quad \Rar \quad \text{Freezing} \label{cAearlylargek} \, , \\
& \cA_i \sim e^{-\epsilon} \qquad & \text{for} \quad \kappa \ll 1 \quad \Rar \quad \text{Freezing} \label{cAearlysmallk} \, ;
\ee
for late times larger than the inverse conductivity $\epsilon\equiv1/(\hat\sigma\eta)\ll1$:
\be
& \cA_i \sim e^{-1/\epsilon} \qquad & \text{for} \quad \kappa \gg 1 \quad \Rar \quad \text{Diffusion} \label{cAlatelargek} \, , \\
& \cA_i \sim e^{-\kappa^2/\epsilon} \qquad & \text{for} \quad \kappa \ll 1 \quad \Rar \quad \text{Freezing} \label{cAlatesmallk} \, ,
\ee
where the last statement is valid for sufficiently small momenta, i.e., large scales.  Thus, diffusion erodes more and more wavelengths as time goes by, if the conductivity remains constant.  Certainly, the fact that it grows can only make things worse.

\paragraph*{Hamiltonian, no instabilities.}
It is instructive to look at the Hamiltonian for this system, since any growing solution for the vector potential would appear there as an instability.  The Euler-Lagrange equation~(\ref{eom}) can directly descend from the classical Hamiltonian
\be\label{ham}
\cH_i = f^2 \left( \pi_i^2 + k^2 \cph_i^2 \right) \, ,
\ee
where $\pi_i = \cA_i'$ and $\cph_i = \cA_i$, and where $\hat\sigma = 2f'/f$.  Thence, this system, in absence of any other interaction term, is positive-definite, and can not give rise to any instability.  There are no growing solutions in the plasma at late times.  This is slightly more general than the result of the previous paragraph, since there is no need to specify the behaviour of the conductivity.

\paragraph*{A lesson learnt.}
The comoving conductivity of the plasma is titanic, see e.g.~Fig.~1 in~\cite{Hollenstein:2012mb}, reaching peaks of $10^{20\div25}/\Mpc$ for redshifts $z\sim0\div10^8$.  Only the largest scales survive the diffusion processes, in absence of other intervening processes such as galactic dynamos of further phase transitions --- which in reality do intervene, but typically at somewhat smaller scales.  At the same time, these large scale Fourier modes are frozen into the plasma and evolve adiabatically with the expansion of the Universe.

Thus, the energy stored in a single mode is also following an adiabatic behaviour, and is comovingly constant.  This implies that the corresponding occupation number does not evolve, and since it is very large now it must have been correspondingly large in the past, at least up until Ohm's law applies, which for large scales is from now backwards until the completion of the reheating process.

\section{Quantum inflationary regime}\label{sec:quant}

\paragraph*{Quantum equations.}
The same action~(\ref{lag}) can be used to describe the inflationary regime, with no current $J_\mu = 0$ as the de Sitter expansion would quickly wash out any pre-existing plasma.  If there is no pre-existing vacuum expectation value for the electromagnetic field at any scale, then the field is quantised canonically: $A_i(\vx,\eta) \rar \varepsilon_i^\lambda(\vk) \cA_i(k,\eta) a^\lambda_\vk + h.c.$, with $[ a^\lambda_\vk, a^{\lambda'\+}_{\vk'} ] = \delta_{\lambda\lambda'} \delta^3 \left( \vk - \vk' \right)$, and where the $\varepsilon$ are polarisation vectors.  The equation of motion for the Fourier modes is simply a free wave equation
\be\label{eomQ}
\cA_i'' + k^2 \cA_i = 0 \, ;
\ee
picking the so-called Bunch-Davies vacuum as initial condition, the solutions are obviously eternally boring and oscillating free waves $\cA_i = \exp(-ik\eta) / \sqrt{2k}$.

The fact that there is a time-dependent background --- the de Sitter expansion --- is unbeknownst to the $\cA_i$, for the reason that~(\ref{lag}) is conformal, that is, does not contain any dimensionful coupling; this combined with the conformal flatness of the metric implies that in the pure de Sitter regime any reference to the scale factor of the Universe $a(\eta)$ is lost.

Notice that this is not the case, for instance, for metric perturbations, massive vector fields, or even massless minimally coupled scalars, for their actions are not conformally invariant.  These examples are briefly touched upon below.

\paragraph*{One occupation...}
The occupation number of the free electromagnetic waves just found is about one, as expected for fields floating around in their vacuum states.  One way of looking at it is to compute the bare power spectrum
\be\label{spectrum}
E_\vk = \frac12 \cP_k \equiv \sum_i \left\{ |\cA_i'|^2 + k^2|\cA_i|^2 \right\} = k \, .
\ee
The spectrum provides the energy stored in each mode, and is integrated over all wavelengths of interest to return the energy density $\rho_\text{em} \sim \int\!\dd k k^3 \cP_k$.  Thence, $n_\vk \sim \cP_k / k \sim 1$ as expected.  Quantum fields in vacuum, in absence of effective interactions with other, classical, fields, happily simmer in their quantum pastoral life.

\paragraph*{...does not lead to many: reheating.}
The de Sitter stage is typically followed by a first order phase transition in which the inflaton field itself ultimately decays into known Standard Model fields and repopulates the Universe.  The dynamics of this epoch are extremely complicated to understand, for it is a non-equilibrium regime, and suddenly finite-temperature too.  A handful reliable statements can however be composed, which will suffice for reaching the argument's conclusion.

The reheating stage is depicted as one where the Universe goes from having one and only one relevant scalar vacuum expectation value, the inflaton one, driving the expansion, to a stage where the Universe is thermal and densely populated by all fields of whichever one's favourite theory is at the energy at which this transition is completed.  Since at the very end the $U(1)$ photon is going to be thrown in the mix, there will be processes which develop an effective conductivity for (the equivalent of) Maxwell's laws.

In particular, this implies that the conductivity, being large at the end of reheating, and zero at the beginning, is doomed to grow during this period.  The build-up of the conductivity in several toy models has been studied for example in~\cite{Son:1996uv,Boyanovsky:1999jh,Giovannini:2000wta,Bassett:2000aw}, where it was indeed found that the rise of the $\hat\sigma$ is very rapid, albeit possibly strongly inhomogeneous.  The conclusion is that, if anything, and in absence of any peculiar coupling for example to the oscillating inflaton, reheating would strongly damp or freeze any electromagnetic field around at that time.

Notice that in principle one needs to solve the out-of-equilibrium, finite-temperature full system of actual (quantum) equations of motion to be able to infer anything with certainty, but, keeping with the spirit of this libel, it is now shown that these bizantinisms are unlikely to bring anything new to the table.  On top of that, imagining any sort of classicalisation during this stage, with the Hubble horizon swallowing more and more wavelengths --- the opposite of what happens during inflation ---, would imply that coherence could build up for modes which stretch well beyond the causal horizon.

\begin{quotation}
\begin{small}
{\bf Interval: renormalisation. }
Another question is that of renormalisation, as in principle the energy density belonging to the electromagnetic field (or any field, for that matter) is formally infinite.  The techniques are now standard background of curved space quantum field theory, and for the electromagnetic field the result is the conformally anomalous term $\rho_\text{em} \sim H^4$ where the constant Hubble parameter is $H \equiv a'/a^2$.  Notice that were all the superhorizon modes have \emph{classical} energy $E_k = aH$ the total \emph{classical} energy would be infinite\footnote{Of course it is not.}: not an extremely smart way of removing infinities.

This fact prompts three quick remarks.  First, this is a purely quantum theoretical result: the conformal anomaly is the quantum vacuum energy density of the field under siege.  The conformal anomaly refers to a renormalised \emph{vacuum} energy density, which by definition is not a highly populated classical state: there is no hint to a transition from quantum to classical in this process.

Second, the modes that cause the infinities and which demand renormalisation are the ultraviolet ones, as in Minkowski space; in fact, the infinities arise in the very same identical way, that is, free fields have infinite energy density because there is an infinity of them --- one at each spacetime point: the only difference is that the subtraction procedure has to be performed in a diffeomorphism invariant way.

Third, inflation should not last forever; it neither extends indefinitely in the past and it obviously decays at a given stage to kick-start the Big Bang.  Since physically (causally) all relevant modes are inside the Hubble horizon $1/aH$ at the onset of inflation, renormalisation should not extend to arbitrary large wavelenghts, as these can never have seen de Sitter approaching.

Furthermore, if there is a coupling and it is sufficiently weak to justify an adiabatic renormalisation procedure, the solutions to the wave equations generalise to
\be\label{adiabatic}
\cA_i^\text{adiabatic} = \frac{1}{\sqrt{2k}} e^{-i\int\omega_k\dd\eta} \, ,
\ee
with the time-dependent frequency $\omega_k$ varying adiabatically: $\omega_k'/\omega_k \ll aH$.  Clearly the particle number is conserved, as it is an adiabatic invariant:
\be\label{adinv}
n_\vk = \frac{E_k}{\omega_k} + \frac12 \sim 1 + \order{\frac{\omega_k'}{aH\omega_k}} \, .
\ee
\end{small}
\end{quotation}

\paragraph*{Quantum-to-classical transition?}
Looking backwards in time the observed magnetic fields approach the end of the reheating stage with a monumental occupation number.  During reheating, at least for the most part, the conductivity grows, so that the magnetic field will at best stay frozen: the occupation number back deep into reheating is still colossal.

But the occupation number at the end of inflation is one; enters reheating, whose detailed quantum field theory, euphemistically speaking, is not yet mastered; what is certain is that that buckets of fields are poured into the primordial brew, and the conductivity grows.  At some point \emph{it might be that} the electromagnetic field begins its pilgrimage to classicalisation.  However, soon the occupation number is going to hit ten, then a hundred, and then a thousand: the field feels already comfortable in its new classical vest; the burden of quantum decoherence is already gone.  Therefore, from this point on the dynamics are very well described by classical equations, and these equations are known:~(\ref{eom}), telling the story of an occupation number which never ever grows again until today.

One may argue that there is a coupling term $J_\mu A^\mu$, or something else more appropriate in the quantum regime, which would do the job of boosting $n_\vk$.  However this, albeit not conformal, only appears \emph{after} the demise of the de Sitter expansion and in the aftermath of inflaton decay, and would incur in precisely the same problem: if the (unknown) effects of such terms --- leaving aside causality issues --- push the occupation number somewhat beyond one, then classical equations apply, and the occupation number on large scales freezes to that value.

\paragraph*{Counterexamples: curvature, magnetogenesis.}
The above ``little no-go'' for vacuum magnetogenesis of course does not apply were the crucial assumption (vacuity, indeed) not to hold.  The first obvious counterexample concerns metric perturbations and the generation, and classicalisation, of curvature perturbations.  For the electromagnetic field itself, any departure from conformal triviality would also do the trick.

For instance, during inflation curvature perturbations are generated because of the time-dependent background; the quantum-to-classical transition can be analysed in depth and is part of the standard lore of inflationary models.  In short, following for example~\cite{Albrecht:1992kf}, one begins with perturbed Einstein equations describing the behaviour of metric perturbations (scalar, vector, and tensor) in a background driven by an almost constant energy density field, the inflaton.  There is a gauge-invariant quantity corresponding to quantised curvature perturbations, oftentimes referred to as ``Mukhanov-Sasaki'' variable, $v(\vx,\eta)$, whose Fourier space Hamiltonian reads
\be\label{vham}
\cH_\vk = p_{-\vk} p_\vk + k^2 v_{-\vk} v_\vk + \frac{z'}{z}\left( p_{-\vk} v_\vk + v_{-\vk} p_\vk \right) \, , \quad z \equiv \frac{a}{H}\left( -\frac{2H'}{3a} \right)^{1/2} \, ,
\ee
with $p_\vk$ the conjugate momentum to $v_\vk$ --- recall that $H'<0$ during slow-roll inflation.  Once again, rewriting in terms of creation and annihilation operators gives a standard $\cH^0_\vk$ plus an interaction term; interaction term which depends entirely on $z(\eta)$, which is in turn connected with the expanding background.  Hence, there is an explicit term in the Euler-Lagrange equation which communicates the effects of the non-trivial background to the quantum Fourier modes.

The equation of motion reads
\be\label{veom}
v_k'' + \left( k^2 - \frac{z''}{z} \right) v_k = 0 \, .
\ee
Choosing again the Bunch-Davies vacuum state as initial condition for all modes, the large scale solution in slow-roll de Sitter expansion is given by
\be
v_k \propto \frac{z}{\sqrt{2k}} \propto \frac{a}{\sqrt{2k}} \qquad \text{for} \quad k^2 \ll \frac{z''}{z} \sim a^2 H^2 \, .
\ee
In this case the particle occupation number is then, from $v_k' \sim aHv_k \gg kv_k$,
\be\label{vocc}
n_k \sim \frac{{v_k'}^2}{k^2} \sim \left(\frac{aH}{k}\right)^2 \gg 1 \, ,
\ee
i.e., for large wavelengths $v_k$ finds its stairway to classical heaven.  It is in this way that curvature perturbations develop a large, quantum first, and then classical, occupation number.

In the context of magnetogenesis the same result can be obtained by explicitly breaking the conformality of the action.  This is typically done using coupling terms such as $f^2F^2$, $gF^*F$, $m^2A^2$ and $j \cdot A$, where $^*F$ is the dual of $F$, and the coupling functions $f(\eta)$, $g(\eta)$, $m(\eta)$, $j(\eta)$ depend on time through the background inflaton.  In fact, the dimensionful mass and current need not depend on time explicitly, but they are usually engineered as such so that at the end of inflation the action reduces to the standard Maxwell one.

The Euler-Lagrange equation displays the appearance of new terms, responsible for the breaking of adiabaticity, and of particle number conservation:
\be\label{eeoomm}
f^2 A_i'' - f^2 \nabla^2 A_i + 2 f f' A_i' + g' \epsilon_{ijk} \d_j A_k + a^2 m^2 A_i + a^2 j_i = 0 \, .
\ee
From here onwards it is simply up to the theorist's taste and inspiration to cook up a theory for these couplings that is efficient and effective.

\section{The end of the occupation}\label{sec:end}

Finally, conclusions.
\begin{itemize}
	\item The occupation number of the free, massless electromagnetic field during de Sitter expansion is of order one for all modes: the field is purely quantum.
	\item In vacuum or in presence of weak (adiabatic) interactions, this number remains constant: the field stays quantum.
	\item At the onset of reheating the field is hence all quantum.
	\item During reheating there might be (not causal?)\ processes which cohere or squeeze the quantum fluctuations.
	\item In any case, once classicalisation begins, the occupation number starts growing from one to ten to a hundred and so on.
	\item At any given desired precision, the field is now classical, and evolves according to classical equations of motion.
	\item The conductivity during reheating builds up rapidly, and forces the magnetic field to either diffuse or freeze: the occupation number decays or, at best, does not change any longer.
	\item After reheating, the occupation number on large scales is again conserved in the plasma due to the pantagruelic conductivity.
	\item Problem: the observed magnetisation of today's large scales implies instead gargantuan occupation numbers: the two giants do not get along with each other.
	\item No occupation number, no party.
\end{itemize}

This ``little no-go'' argumentation is entirely based on one assumption: the vacuity of the initial electromagnetic field, and its persistence in the inflating regime.  For most magnetogenesis models this assumption does not apply, because there normally is a mechanism which inflates the electromagnetic occupation number together with the Universe.  Lacking such additional gear, inflation alone won't do.

\section*{Acknowledgements}
FU and PT are supported by IISN project No.~4.4502.13 and Belgian Science Policy under IAP VII/37.

\vspace{0.6in}
\begin{flushright}
\emph{Listen to me\\
Listen to me\\
Listen to me\\
Listen to me\\}
R.E.M., Welcome to the occupation
\end{flushright}

\newpage

\appendix
\section*{Appendix}

\paragraph*{Relation to squeezing.}
Squeezed states are often used in discussing the issue of classicalisation.  Although squeezed states can be shown to not be a typical classical state in the sense minimal Heisenberg uncertainty, they can also be shown to be extremely classical in a WKB sense --- that is, adiabatically.  For example, it is well known that metric perturbations, and in turn curvature perturbations, squeeze dramatically once they leave the horizon during inflation.  This can be seen again by solving the Euler-Lagrange equation for $v_k$:~(\ref{veom}).  Using the interaction Hamiltonian to define the time-evolution operators in the Schr\'odinger representation one defines a squeezing factor $R_\vk$ which essentially tells how much the phase space of a given state ``squeezes'' along some given axis.  In the curvature perturbation case this is shown to obey
\be\label{squeeze}
R_k = \text{arcsinh}\frac{1}{2k\eta} \gg 1 \qquad \text{for} \quad k\eta \ll 1 \, . \nn
\ee
Moreover, there is a direct relation between the squeezing and the occupation number:
\be\label{occupyandsqueeze}
n_\vk = \sinh^2 R_\vk \rar \left(\frac{aH}{k}\right)^2 \, , \nn
\ee
as~(\ref{vocc}) also figured.  The corresponding situation for the free electromagnetic field is that the squeezing factor is of order one for order one occupation number: no squeezing at all, thence no WKB classicality.

\begin{quotation}
\begin{small}
{\bf Addendum.}
In the process of renormalising a vacuum expectation value, all sorts of quirky things can happen: negative energies, rational numbers of spacetime dimensions, etc.  However, at the end, the final result should be physically sensible.  There does not seem to be a fundamental problem with negative power spectra as long as these turn to positive-define once physical quantities are computed.  However, \emph{physical} negative energy densities are only acceptable, as in the Casimir effect, when part of a system whose \emph{total} energy density is well defined.  If the result of some renormalisation procedure for the Universe's vacuum energy returns a negative value, being there no \emph{outer environment}, said procedure should at least be regarded with sheer suspicion.
\end{small}
\end{quotation}


\begin{thebibliography}{99}

\bibitem{Kronberg:1993vk}
  P.~P.~Kronberg,
  Rept.\ Prog.\ Phys.\  {\bf 57} (1994) 325.

\bibitem{Grasso:2000wj}
  D.~Grasso and H.~R.~Rubinstein,
  Phys.\ Rept.\  {\bf 348} (2001) 163
  [astro-ph/0009061].

\bibitem{Kulsrud:2007an}
  R.~M.~Kulsrud and E.~G.~Zweibel,
  Rept.\ Prog.\ Phys.\  {\bf 71} (2008) 0046091
  [arXiv:0707.2783 [astro-ph]].

\bibitem{Kandus:2010nw}
  A.~Kandus, K.~E.~Kunze and C.~G.~Tsagas,
  Phys.\ Rept.\  {\bf 505} (2011) 1
  [arXiv:1007.3891 [astro-ph.CO]].

\bibitem{Widrow:2011hs}
  L.~M.~Widrow, D.~Ryu, D.~R.~G.~Schleicher, K.~Subramanian, C.~G.~Tsagas and R.~A.~Treumann,
  Space Sci.\ Rev.\  {\bf 166} (2012) 37
  [arXiv:1109.4052 [astro-ph.CO]].

\bibitem{Durrer:2013pga}
  R.~Durrer and A.~Neronov,
  arXiv:1303.7121 [astro-ph.CO].

\bibitem{Albrecht:1992kf}
  A.~Albrecht, P.~Ferreira, M.~Joyce and T.~Prokopec,
  Phys.\ Rev.\ D {\bf 50} (1994) 4807
  [astro-ph/9303001].

\bibitem{Finelli:2007fr}
  F.~Finelli, G.~Marozzi, G.~P.~Vacca and G.~Venturi,
  Phys.\ Rev.\ D {\bf 76} (2007) 103528
  [arXiv:0707.1416 [hep-th]].

\bibitem{Durrer:2009ii}
  R.~Durrer, G.~Marozzi and M.~Rinaldi,
  Phys.\ Rev.\ D {\bf 80} (2009) 065024
  [arXiv:0906.4772 [astro-ph.CO]].

\bibitem{Marozzi:2011da}
  G.~Marozzi, M.~Rinaldi and R.~Durrer,
  Phys.\ Rev.\ D {\bf 83} (2011) 105017
  [arXiv:1102.2206 [astro-ph.CO]].

\bibitem{Campanelli:2013mea}
  L.~Campanelli,
  arXiv:1304.6534 [astro-ph.CO].

\bibitem{Durrer:2013xla}
  R.~Durrer, G.~Marozzi and M.~Rinaldi,
  arXiv:1305.3192 [astro-ph.CO].

\bibitem{Campanelli:2013fqa}
  L.~Campanelli,
  arXiv:1305.7062 [astro-ph.CO].

\bibitem{Brandenburg:2004jv}
  A.~Brandenburg and K.~Subramanian,
  Phys.\ Rept.\  {\bf 417} (2005) 1
  [astro-ph/0405052].

\bibitem{Hollenstein:2012mb}
  L.~Hollenstein, R.~K.~Jain and F.~R.~Urban,
  JCAP {\bf 1301} (2013) 013
  [arXiv:1208.6547 [astro-ph.CO]].

\bibitem{Son:1996uv}
  D.~T.~Son,
  Phys.\ Rev.\ D {\bf 54} (1996) 3745
  [hep-ph/9604340].

\bibitem{Boyanovsky:1999jh}
  D.~Boyanovsky, H.~J.~de Vega and M.~Simionato,
  Phys.\ Rev.\ D {\bf 61} (2000) 085007
  [hep-ph/9909259].

\bibitem{Giovannini:2000wta}
  M.~Giovannini and M.~E.~Shaposhnikov,
  hep-ph/0011105.

\bibitem{Bassett:2000aw}
  B.~A.~Bassett, G.~Pollifrone, S.~Tsujikawa and F.~Viniegra,
  Phys.\ Rev.\ D {\bf 63} (2001) 103515
  [astro-ph/0010628].

\end{thebibliography}
\end{document}